\documentclass[%
reprint,
showpacs,preprintnumbers,
 amsmath,amssymb,
 aps,
]{revtex4-1}

\usepackage{graphicx}
\usepackage{dcolumn}
\usepackage{bm}
\usepackage{subfigure} 
\usepackage{color,soul}
\usepackage{ulem}
\usepackage{booktabs}

\begin{document}


\title{Topography- and topology-driven spreading of non-Newtonian power-law liquids on a flat and a spherical substrate
}

\author{Masao Iwamatsu}
\email{iwamatsu@ph.ns.tcu.ac.jp}
\affiliation{%
Department of Physics, Faculty of Liberal Arts and Sciences, Tokyo City University, Setagaya-ku, Tokyo 158-8557, Japan
}%



\date{\today}

\begin{abstract}
The spreading of a cap-shaped spherical droplet of non-Newtonian power-law liquids on a flat and a spherical rough and textured substrate is theoretically studied in the capillary-controlled spreading regime.  A droplet whose scale is much larger than that of the roughness of substrate is considered.  The equilibrium contact angle on a rough substrate is modeled by the Wenzel and the Cassie-Baxter model.   Only the viscous energy dissipation within the droplet volume is considered, and that within the texture of substrate by imbibition is neglected.  Then, the energy balance approach is adopted to derive the evolution equation of the contact angle. When the equilibrium contact angle vanishes, the relaxation of dynamic contact angle $\theta$ of a droplet obeys a power law decay $\theta \sim t^{-\alpha}$ except for the Newtonian and the non-Newtonian shear-thinning liquid of the Wenzel model on a spherical substrate.  The spreading exponent $\alpha$ of the non-Newtonian shear-thickening liquid of the Wenzel model on a spherical substrate is larger than others.  The relaxation of the Newtonian liquid of the Wenzel model on a spherical substrate is even faster showing the exponential relaxation.  The relaxation of the non-Newtonian  shear-thinning liquid of Wenzel model on a spherical substrate is fastest and finishes within a finite time.  Thus, the topography (roughness) and the topology (flat to spherical) of substrate accelerate the spreading of droplet.
\end{abstract}

\maketitle

\section{Introduction}
The spreading of a liquid droplet on a solid substrate has been studied for many years, because it plays fundamental roles in many natural phenomena and industrial applications~\cite{deGennes1985,Daniel2006,Bonn2009}.  Even though the spreading of a liquid droplet on a solid substrate is a complicated phenomena where many factors come into play, the time evolution of spreading on a flat~\cite{Hoffman1975,Voinov1976,Tanner1979,Hervet1984,deGennes1985,Seaver1994,deRuijter2000,Daniel2006} as well as on a spherical~\cite{Iwamatsu2017,Iwamatsu2017b} solid substrate is usually described by amazingly simple universal power laws.  The most well-known law called Tanner's law describes the spreading of a small non-volatile droplet of Newtonian liquids on a completely wettable flat smooth substrate.  This law has been derived theoretically from several different approaches~\cite{Voinov1976,Tanner1979,Hervet1984} and confirmed experimentally~\cite{Tanner1979,deRuijter1999,Rafai2004}. However, most of theoretical as well as experimental works are confined to the spreading on a smooth substrate.

The wetting of an artificially or a naturally rough and textured substrate~\cite{Shuttleworth1948,Bormashenko2015} is different from the wetting of smooth substrates.  Usually, the wetting of rough substrates is modeled by two macroscopic models called Wenzel model~\cite{Wenzel1936} and Cassie-Baxter model~\cite{Cassie1944}.  These models predict that the equilibrium contact angle is not given by the Young's~\cite{Young1805} formula from the composition of the substrate and the liquid, but is modified by the roughness of the substrate~\cite{Wenzel1936,Cassie1944}.  Although the validity of the Wenzel model and the Cassie-Baxter model was questioned~\cite{McHale2007,Gao2007,Iwamatsu2006}, they are still applicable to real wetting within a limited condition~\cite{McHale2007}.

In the present study, we integrate our knowledge of the spreading of Newtonian~\cite{deGennes1985,Daniel2006,Bonn2009,Iwamatsu2017} as well as non-Newtonian~\cite{Wang2007,Liang2009,Dandapat2010,Liang2012,Iwamatsu2017b} liquid droplets on a smooth substrate, and consider the spreading on a flat and a spherical rough substrate.  In particular, we pay attention to the spreading towards the complete wetting state of zero contact angle.  The complete wetting can be realized by an incompletely wettable substrate with non-zero Young's contact angle by the roughness in the Wenzel model~\cite{Wenzel1936,Shibuich1996,Quere2002,Bico2002,Zhang2014}, which has been attracted much attention as the super-wetting, super-spreading and antifogging materials~\cite{Shibuich1996,Drelich2010,Bahners2013,Fujima2014,Zhang2014,Wang2015}.

More specifically, we consider the problem of spreading of a cap-shaped spherical droplet of non-volatile non-Newtonian liquids gently placed on the top of a flat and of a spherical rough substrate using the energy-balance approach~\cite{Hervet1984,deGennes1985,Daniel2006} which can be easily extended to non-Newtonian liquid~\cite{Liang2009,Liang2012} and to spherical geometry~\cite{Iwamatsu2017b}.  The problem of spreading on a flat rough substrate using the Wenzel model has already been considered theoretically and has been tested experimentally by McHale {\it et al.}~\cite{McHale2004,McHale2009}. 
They termed the spreading of a droplet described by the Wenzel model on a flat rough substrate {\it topography driven} spreading because the complete wetting is achieved by the topography (roughness) of the substrate. However, they studied only Newtonian liquids.  Later, Singh and Dandapat~\cite{Singh2013} studied the same problem for non-Newtonian liquids including the line tension effect.  Those previous works~\cite{McHale2004,McHale2009,Singh2013}, however, studied the problem of spreading only on a flat substrate.

Here, we consider the same problem of spreading of a non-Newtonian liquid droplet not only on a flat but also on a spherical rough substrate.  Then, the spreading will be affected not only by the {\it topography} (roughness) but also by the (flat and spherical) {\it topology} of the substrates.  Therefore, we study the {\it topography and topology driven} spreading using the energy-balance approach~\cite{Hervet1984,deGennes1985,Daniel2006,Liang2012,Iwamatsu2017b,McHale2009,Singh2013}.  We assume that the roughness is mild and the size of roughness is much smaller than the droplet size so that the liquid volume invading into the texture of the substrate is negligible. We further assume that the energy dissipation due to the imbibition~\cite{Ishino2007,Haidara2008,Grewal2015} within the texture of substrate is negligible.  Hence, the roughness of the substrate affects only the driving force (capillary force) of spreading through the equilibrium contact angle that is modified by the roughness.

This paper is organized as follows. In Sec. II, we summarize the essence of the Wenzel~\cite{Wenzel1936} and the Cassie-Baxter~\cite{Cassie1944} model of wetting on rough substrate. In Sec. III, we present main results of topography and topology driven and enhanced spreading, in particular, the scaling laws of spreading on a flat and a spherical rough substrate.  Finally, in Sec. IV, we conclude by emphasizing the most important implication of our results.

\section{\label{sec:sec2}Equilibrium contact angle of Wenzel and Cassie droplets}

\begin{figure}[htbp]
\begin{center}
\includegraphics[width=0.80\linewidth]{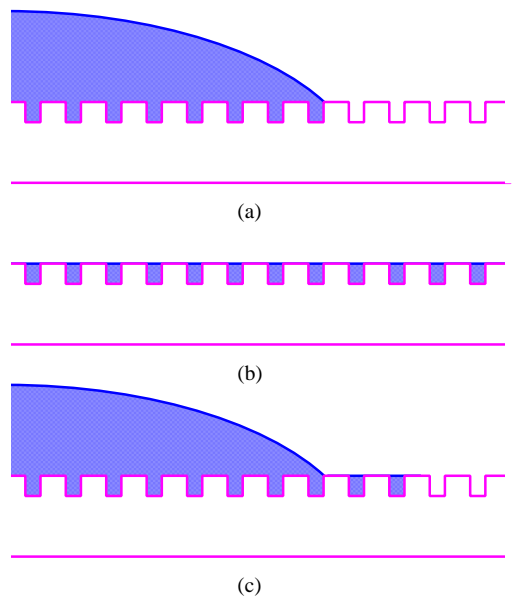}
\caption{
(a) The Wenzel droplet which invades the roughness of the substrate underneath the contact base of the droplet.  (b) The hemi-wicked state where the liquid completely invade the texture of the substrate, and the droplet volume above the substrate disappears.  (c) The Cassie droplet, which spreads on the composite substrate composed of the solid substrate and the liquid precursor film ahead of the contact line of the droplet.
}
\label{fig:T1}
\end{center}
\end{figure}

When the substrate is rough and textured, the liquid of droplet invades the texture and completely wets the inside of the substrate beneath the droplet (Fig.~\ref{fig:T1}(a)).  This model is called Wenzel model, and the equilibrium contact angle $\theta_{\rm e}$ is given by the Wenzel formula $\theta_{\rm e}=\theta_{\rm W}$ defined by~\cite{Wenzel1936,Bico2002}
\begin{equation}
\cos\theta_{\rm W}=r_{\rm s}\cos\theta_{\rm Y},
\label{eq:T1}
\end{equation}
with the roughness of the substrate $r_{\rm s}$ defined by~\cite{McHale2009} 
\begin{equation}
r_{\rm s}=\frac{\Delta A_{\rm wetted}}{\Delta A_{\rm p}},
\label{eq:T2}
\end{equation}
where $\Delta A_{\rm wetted}$ is a small change in wetted area that is sampled by the advanced three-phase contact line and $\Delta A_{\rm p}$ is the planar projection of that change.  Strictly speaking, the roughness $r_{\rm s}$ depends on the position of contact line~\cite{Iwamatsu2006,McHale2009}.  However, we will consider a large droplet whose scale is much larger than the scale of roughness so that the roughness $r_{\rm s}$ can be regarded as uniform and constant.  The substrate composition and liquid is characterized by the Young's contact angle $\theta_{\rm Y}$, which is related to the liquid-vapor (LV), liquid-solid (LS), and solid-vapor (SV) interface energies $\sigma_{\rm LV}$, $\sigma_{\rm LS}$, and $\sigma_{\rm SV}$ through
\begin{equation}
\cos\theta_{\rm Y}=\frac{\sigma_{\rm SL}-\sigma_{\rm SV}}{\sigma_{\rm LV}}.
\label{eq:T3}
\end{equation}
We will call the droplet whose contact angle is dictated by the Wenzel formula in Eq.~(\ref{eq:T1}) "Wenzel droplet" in this paper.

Equation (\ref{eq:T1}) suggests that an incompletely wettable substrate with $\theta_{\rm Y}<90^{\circ}$ shows a lower equilibrium contact angle than the Young's contact angle $\theta_{\rm W}<\theta_{\rm Y}$ since $r_{\rm s}>1$.  The roughness improves the wettablility  (hydrophilicity) of the substrate, and the equilibrium contact angle can be as small as $\theta_{\rm W}=0^{\circ}$. Then, the substrate becomes completely wettable (superhydrophilic) when
\begin{equation}
r_{\rm s}>\frac{1}{\cos\theta_{\rm Y}}.
\label{eq:T4}
\end{equation}
Therefore, the {\it topography} (roughness) driven complete wetting ($\theta_{\rm W}=0^{\circ}$) super-hydrophilic substrate can be manufactured~\cite{Zhang2014} using an incompletely wettable materials with $\theta_{\rm Y}>0^{\circ}$ when Eq.~(\ref{eq:T4}) is satisfied.  In this paper, we will use the terminology "wettable", "hydrophilic", and "hydrophobic" interchangeably though the liquid is not necessarily water.

A complete invasion of liquid into the texture of substrate called hemi-wicking~\cite{Quere2002,Bico2002} (Fig.~\ref{fig:T1}(b)) is thermodynamically stabler than the Wenzel model when
\begin{equation}
\theta_{\rm Y}<\theta_{\rm c},
\label{eq:T5}
\end{equation}
where
\begin{equation}
\cos\theta_{\rm c}=\frac{1-\phi_{\rm s}}{r_{\rm s}-\phi_{\rm s}}.
\label{eq:T6}
\end{equation}
The fraction $\phi_{\rm s}<1$ is defined through
\begin{equation}
\phi_{\rm s}=\frac{\Delta A_{\rm non-wetted}}{\Delta A_{\rm p}},
\label{eq:T7}
\end{equation}
where $\Delta A_{\rm non-wetted}$ is a small increase in flat-top dry area (Fig.~\ref{fig:T1}(b)) that is sampled by the advanced three-phase contact line.  Then, the condition of hemi-wicked state given by Eq.~(\ref{eq:T5}) is also written as
\begin{equation}
\phi_{\rm s}>\frac{1-\cos\theta_{\rm W}}{1-\cos\theta_{\rm Y}}.
\label{eq:T8}
\end{equation}
Therefore, a sufficient amount of dry surface area which is exposed to the vapor phase is necessary for hemi-wicking (complete imbibition) to occur.  For a topography driven complete wetting (superhydrophilic) substrate with $\theta_{\rm W}=0^{\circ}$, the hemi-wicking condition Eq.~(\ref{eq:T8}) becomes $\phi_{\rm s}>0$.  Then, the hemi-wicked state is always stabler than the Wenzel state.   

Note that the complete wetting state with $\theta=0^{\circ}$ is different from the hemi-wicked state.  In the former case, the flat-top area of the rough substrate is covered by a thin layer of wetting film. By contrast, the flat-top area remains dry in the hemi-wicked state.  Therefore, thin wetting film spreads over the whole area of completely wettable rough substrate as the contact angle approaches zero ($\theta\rightarrow 0^{\circ}$).  In this limit of thin film, the spreading and the imbibition cannot be separated and the liquid volume invading into the texture cannot be neglected.  We will not consider this thin-film limit, but will focus on the spreading process towards the complete wetting state when the liquid volume in the texture is still negligible.    

When the condition given by Eq.~(\ref{eq:T5}) is satisfied, the droplet spreading and the hemi-wicking can occur simultaneously.  In such a case, the droplet spread on a substrate whose roughness is already filled by the liquid precursor film~\cite{Neogi1983} because the time ($t$) scale of imbibition~\cite{Washburn1921} $\sqrt{t}$ would be faster than the time scale of spreading.  Then, the droplet spreads on a composite substrate made by the solid and the precursor liquid ahead of the spreading front (Fig.~\ref{fig:T1}(c)).  The equilibrium contact angle $\theta_{\rm e}$ is given by the so-called Cassie-Baxter formula ($\theta_{\rm e}=\theta_{\rm CB}$)~\cite{Quere2002,Bico2002,McHale2009}:
\begin{equation}
\cos\theta_{\rm CB}=\phi_{\rm s}\cos\theta_{\rm Y}+\left(1-\phi_{\rm s}\right).
\label{eq:T9}
\end{equation}
We will call this droplet whose equilibrium contact angle is determined from the Cassie-Baxter formula in Eq.~(\ref{eq:T9}) "Cassie" droplet.  

\begin{figure}[htbp]
\begin{center}
\includegraphics[width=1.0\linewidth]{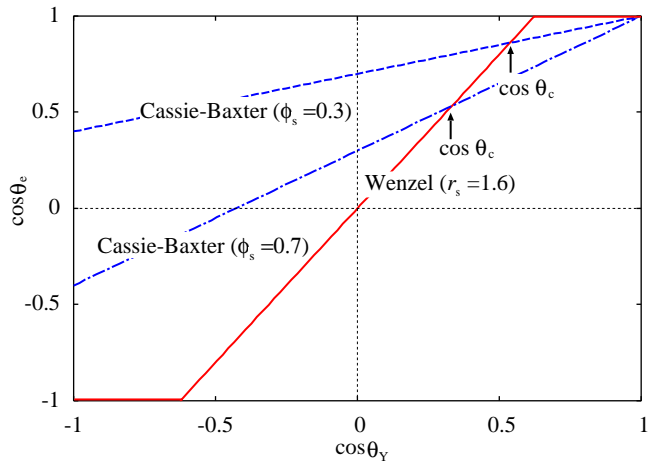}
\caption{
The cosine of the equilibrium contact angle $\cos\theta_{\rm W}$ ($\theta_{\rm e}=\theta_{\rm W}$) of the Wenzel droplet with $r_{\rm s}=1.6$ and $\cos\theta_{\rm CB}$ ($\theta_{\rm e}=\theta_{\rm CB}$) of the Cassie droplet with $\phi_{\rm s}=0.3$ and $\phi_{\rm s}=0.7$ calculated from Eqs.~(\ref{eq:T1}) and (\ref{eq:T9}) as a function of the cosine of Young's contact angle $\cos\theta_{\rm Y}$.  The Cassie droplet will be more stable than Wenzel droplet when $\cos\theta_{\rm Y}>\cos\theta_{\rm c}$.   
 }
\label{fig:T2}
\end{center}
\end{figure}

In Fig.~\ref{fig:T2}, we show the cosine of the equilibrium contact angle $
\cos\theta_{\rm W}$ of the Wenzel droplet and $\cos\theta_{\rm CB}$ of the Cassie droplet versus the cosine of the Young's contact angle $\cos\theta_{\rm Y}$ when $r_{\rm s}=1.6$, and $\phi_{\rm s}=0.3$ and $0.7$.  The complete wetting (superhydrophilic) state  ($\theta_{\rm W}=0^{\circ}$ and $\cos\theta_{\rm W}=1$) appears in the Wenzel state when $\cos\theta_{\rm Y}>1/r_{\rm s}=0.625$.  However, Equation~(\ref{eq:T5}) suggests that this completely wettable Wenzel state may not be realized. Rather, the Cassie state may appear~\cite{Quere2002,Bico2002}.  This scenario is consistent to the beautiful experiment by Shibuich et al.~\cite{Shibuich1996}.  In the Wenzel state, the substrate becomes more hydrophilic ($\cos\theta_{\rm W}>\cos\theta_{\rm Y}>0$) when $\theta_{\rm Y}<90^{\circ}$, and it becomes more hydrophobic ($\cos\theta_{\rm W}<\cos\theta_{\rm Y}<0$) when $\theta_{\rm Y}>90^{\circ}$.

For the Cassie droplet on a hemi-wicked substrate (Fig.~\ref{fig:T2}), the complete wetting  (superhydrophilic) state ($\theta_{\rm CB}=0^{\circ}$) appears only when the composition of the substrate and liquid is also completely wettable ($\theta_{\rm Y}=0^{\circ}$).  However, the substrate becomes more wettable overall ($\cos\theta_{\rm CB}>\cos\theta_{\rm Y}$) by the roughness.  In fact, even a hydrophobic substrate with $\cos\theta_{\rm Y}<0$ turns to a hydrophilic substrate ($\cos\theta_{\rm CB}>0$) when $\phi_{\rm s}<0.5$ (Fig.~\ref{fig:T2}).  Then, the substrate is always hydrophilic irrespective of the material used for the substrate and the liquid.  

It is possible to consider a mixed wetting state which is intermediate between Wenzel state and Cassie-Baxter state~\cite{Marmur2003,Bormashenko2011}, whose equilibrium contact angle $\theta_{\rm mix}$ is given by 
\begin{equation}
\cos\theta_{\rm mix}=r_{\rm f}f\cos\theta_{\rm Y}+f-1,
\label{eq:T10}
\end{equation}
where $f$ is the fraction of the projected area of the solid surface that is wetted by the liquid, and $r_{\rm f}$ is the roughness ratio of the wet area.  When $f=1$ and $r_{\rm f}=r_{\rm s}$, Eq.~(\ref{eq:T10}) becomes Eq.~(\ref{eq:T1}) of the Wenzel state.  This mixed wetting state has been discussed in detail by Marmur~\cite{Marmur2003}.  But, we will only consider the Wenzel state in Eq.~(\ref{eq:T1}) and the Cassie penetrating (impregnating) state~\cite{Bormashenko2015} in Eq.~(\ref{eq:T9}).

\section{\label{sec:sec3}Spreading law of non-Newtonian liquids on a flat and a spherical rough substrate}

In the following, we will consider the spreading of a cap-shaped spherical droplet of non-Newtonian liquids on a flat~\cite{Liang2012} and a spherical substrate~\cite{Iwamatsu2017b} using these Wenzel and Cassie droplet models.  We assume that the droplet volume is sufficiently large so that the liquid volume invading into the substrate is negligible.  To make our discussion as general as possible, we consider non-Newtonian power-law liquids~\cite{Wang2007,Liang2009,Dandapat2010,Liang2012,Iwamatsu2017b}.  The apparent viscosity $\mu$ depends on the shear rate $\dot{\gamma}$ through
\begin{equation}
\mu = \kappa \dot{\gamma}^{n-1},
\label{eq:T11}
\end{equation}
where $\kappa$ is a consistency coefficient~\cite{Wang2007,Liang2012}. The power-law exponent $n$ characterizes the non-Newtonian liquids.  When $n>1$, the liquid is called shear thickening.  When $n<1$, it is called shear shinning.  The usual Newtonian liquids correspond to $n=1$.

\subsection{Spreading on a planar rough substrate}

\begin{figure}[htbp]
\begin{center}
\includegraphics[width=1.0\linewidth]{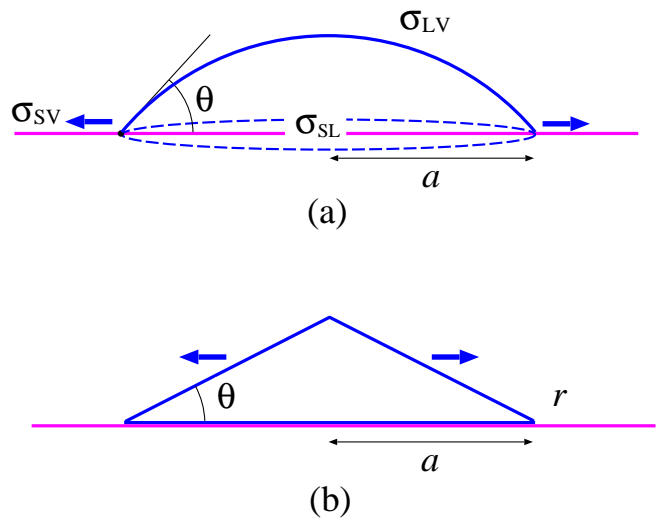}
\caption{
(a) A droplet on a flat substrate spreading toward the complete-wetting state.  The base radius $a$ and the three-phase contact line expand towards infinity.  (b) The cone-shaped model of spreading droplet on a flat substrate.  The meniscus is approximated by a wedge.
 }
\label{fig:T3}
\end{center}
\end{figure}

A droplet placed gently on a solid substrate spreads until its dynamic contact angle $\theta$ reaches the stable equilibrium value  $\theta_{\rm e}$, which is given either by $\theta_{\rm W}$ defined by Eq.~(\ref{eq:T1}) or $\theta_{\rm CB}$ defined by Eq.~(\ref{eq:T9}) on a rough substrate. The driving force $f_{\rm L}$ for this spreading is the out-of-balance component of the capillary force parallel to the surface acting along the contact line L, and is given by
\begin{equation}
f_{\rm L}=\sigma_{\rm LV}\left(\cos\theta_{\rm e}-\cos\theta\right) - \frac{\tau}{a},
\label{eq:T12}
\end{equation}
where $a$ is the base radius (Fig.~\ref{fig:T3}), and $\tau$ is the line tension~\cite{Navascues1981}.  Because the nature of the line tension on rough substrates is largely unknown, and its inclusion in Eq.~(\ref{eq:T12}) would be highly debatable, we will neglect the last term of line tension contribution in Eq.~(\ref{eq:T12}).  Also, since we consider a macroscopic droplet much larger than the scale of roughness, we can safely neglect the line tension in the following discussions as the line tension is believed to become important only for a nano-scale droplet~\cite{Schimmele2007,Law2017}.

Since we neglect the liquid volume invading into the rough substrate, we consider only the viscous dissipation $\dot{\Sigma}_{\rm drop}$ within the cap-shaped spherical droplet, which can be approximated by the viscous dissipation within the cone-shaped droplet with a wedge-shaped meniscus shown in Fig.~\ref{fig:T3}(b).  The result for the cone-shaped droplet is given by~\cite{Liang2012}
\begin{equation}
\dot{\Sigma}_{\rm drop}=2\pi k\left(\frac{8n+4}{3n}\right)^{n}a^{2-n}\frac{\kappa U^{n+1}}{\theta^{n}},
\label{eq:T13}
\end{equation}
where $U$ is the spreading velocity and the pre-factor $k$ depends on the cut-off length of the wedge-shaped meniscus.

Since Eq.~(\ref{eq:T13}) is based on the wedge approximation~\cite{Liang2012}, its validity might be questionable when the dynamic contact angle $\theta$ approaches $0^{\circ}$ and the droplet becomes flat. Then, the droplet near the contact line looks more like a thin film or a slab rather than a wedge. However, an equation similar to Eq.~(\ref{eq:T13}) can be derived by simplifying a spherical droplet by a thin film or a disk~\cite{Daniel2006,deRuijter1999,Seaver1994} instead of a cone for the Newtonian liquids. Therefore, Eq.~(\ref{eq:T13}) for the non-Newtonian liquids would be applicable as well, even when the contact angle $\theta$ approaches $0^{\circ}$.

The energy-balance condition at the contact line with radius $a$
\begin{equation}
2\pi a f_{\rm L}U=\dot{\Sigma}_{\rm drop}
\label{eq:T14}
\end{equation}
is given by
\begin{equation}
\theta^{n}\left(\cos\theta_{\rm e}-\cos\theta\right)
=k\left(\frac{8n+4}{3n}\right)^{n}a^{1-n}\frac{\kappa U^{n}}{\sigma_{\rm LV}}.
\label{eq:T15}
\end{equation}
This energy-balance approach in Eq.~(\ref{eq:T14}) is valid only when the spreading velocity $U$ is low and the viscous length scale is larger than other length scales~\cite{Bonn2009} so that the inertial effect can be neglected.  The base radius $a$ is related to the droplet volume $V_{0}$ through
\begin{equation}
V_{0}=\frac{\pi}{4}\theta a^{3},
\label{eq:T16}
\end{equation}
for a spherical-cap droplet with a small contact angle $\theta$ on a flat substrate.  Then the energy-balance condition in Eq.~(\ref{eq:T15}) is written as
\begin{eqnarray}
&&\theta^{\frac{2n+1}{3}}\left(\cos\theta_{\rm e}-\cos\theta\right)
\nonumber \\
&&=2^{\frac{4n+2}{3}}k\left(\frac{2n+1}{3n}\right)^{n}\left(\frac{V_{0}}{\pi}\right)^{\frac{1-n}{3}}\frac{\kappa U^{n}}{\sigma_{\rm LV}},
\label{eq:T17}
\end{eqnarray}
which reduces to the universal law~\cite{Hoffman1975,Voinov1976,Tanner1979,Seaver1994}
\begin{equation}
\theta^{3} \propto {\rm Ca},
\label{eq:T18}
\end{equation}
for the complete-wetting ($\theta_{\rm e}=\theta_{\rm Y}=0^{\circ}$) Newtonian liquids ($n=1$) on a smooth surface, where ${\rm Ca}=\kappa U/\sigma_{\rm LV}$ is the capillary number.

Since the equilibrium contact angle $\theta_{\rm e}$ is given either by $\theta_{\rm W}$ in Eq.~(\ref{eq:T1}) for the Wenzel droplet or by $\theta_{\rm CB}$ in Eq.~(\ref{eq:T9}) for the Cassie droplet, the driving force in Eq.~(\ref{eq:T12}) for the Cassie droplet is given by
\begin{equation}
f_{\rm L}=\sigma_{\rm LV}\left[\phi_{\rm s}\cos\theta_{\rm Y}+\left(1-\phi_{\rm s}\right)-\cos\theta\right].
\label{eq:T19}
\end{equation}
Then, Eq.~(\ref{eq:T17}) is further simplified to
\begin{eqnarray}
&&\frac{1}{2}\theta^{\frac{2n+1}{3}}\left(\theta^{2}-\phi_{\rm s}\theta_{\rm Y}^{2}\right)
\nonumber \\
&&=2^{\frac{4n+2}{3}}k\left(\frac{2n+1}{3n}\right)^{n}\left(\frac{V_{0}}{\pi}\right)^{\frac{1-n}{3}}\frac{\kappa U^{n}}{\sigma_{\rm LV}},
\label{eq:T20}
\end{eqnarray}
for the Cassie droplet when the dynamic contact angle $\theta$ and the Young's contact angle $\theta_{\rm Y}$ are small.  The complete wetting state is achieved only when $\theta_{\rm Y}=0^{\circ}$ for the Cassei droplet, and Eq.~(\ref{eq:T20}) reduces to 
\begin{equation}
\theta^{\frac{2n+7}{3}}\propto U^{n},
\label{eq:T21}
\end{equation}
which is the same~\cite{Liang2012,Wang2007} as that derived for a droplet on a smooth flat substrate.  Equation (\ref{eq:T21}) also reduces to Eq.~(\ref{eq:T18}) for Newtonian liquids when $n=1$.

By contrast, the driving force for the Wenzel droplet is given by
\begin{equation}
f_{\rm L}=\sigma_{\rm LV}\left(r_{\rm s}\cos\theta_{\rm Y}-\cos\theta\right),
\label{eq:T22}
\end{equation}
and Eq.~(\ref{eq:T17}) is simplified to
\begin{eqnarray}
&&\theta^{\frac{2n+1}{3}}\left[\left(r_{\rm s}\cos\theta_{\rm Y}-1\right)+\frac{1}{2}\theta^{2}\right]
\nonumber \\
&&=2^{\frac{4n+2}{3}} k\left(\frac{2n+1}{3n}\right)^{n}\left(\frac{V_{0}}{\pi}\right)^{\frac{1-n}{3}}\frac{\kappa U^{n}}{\sigma_{\rm LV}},
\label{eq:T23}
\end{eqnarray}
when the dynamic contact angle $\theta$ becomes small. In this case, the complete wetting state $\theta_{\rm W}=0^{\circ}$ is possible even when $\theta_{\rm Y}>0^{\circ}$ as long as Eq.~(\ref{eq:T4}) is satisfied.  Although this Wenzel complete wetting state with $\theta_{\rm Y}>0^{\circ}$ is thermodynamically less stable and the Cassie droplet state would appear, we will continue to consider the situation when $\theta_{\rm W}=0^{\circ}$ and $\theta_{\rm Y}\ge 0^{\circ}$ to make our discussion more general.  

Then, Eq.~(\ref{eq:T23}) for the Wenzel droplet reduces to
\begin{equation}
\theta^{\frac{2n+1}{3}} \propto U^{n},
\label{eq:T24}
\end{equation}
which further reduces to
\begin{equation}
\theta \propto {\rm Ca},
\label{eq:T25}
\end{equation}
for Newtonian liquids with $n=1$, which is different from  Eq.~(\ref{eq:T18}) for the Cassie droplet and the free droplet on a smooth substrate~\cite{Hoffman1975,Voinov1976,Tanner1979,Seaver1994}.  Equation (\ref{eq:T24}) for the Wenzel droplet is also different from Eq.~(\ref{eq:T21}) for the Cassie droplet.  Therefore, the scaling laws for the Wenzel droplet will be different from those for the Cassie droplet. Since the driving force of spreading in Eq.~(\ref{eq:T22}) becomes constant $f_{\rm L}\propto\left(r_{\rm s}\cos\theta_{\rm Y}-1\right)$ for the Wenzel droplet when $\theta\rightarrow 0^{\circ}$ while the spreading force for the Cassie droplet in Eq.~(\ref{eq:T19}) vanishes when $\theta_{\rm Y}=0^{\circ}$ and  $\theta\rightarrow 0^{\circ}$, the Wenzel droplet is always accelerated and spreads faster than the Cassie droplet.  These facts were already pointed out on a flat rough substrate by McGraw et al.~\cite{McHale2004,McHale2009} for the Newtonian liquids with $n=1$.

On a completely wettable substrate with $\theta_{\rm Y}=0^{\circ}$, the evolution equation Eq. (\ref{eq:T20}) for the Cassie droplet becomes the same as that for the free droplet on a completely wettable smooth substrate~\cite{Liang2012}.  Since the spreading velocity is given by
\begin{equation}
U=\frac{da}{dt}=-\frac{4^{\frac{1}{3}}}{3\theta^{\frac{4}{3}}}\left(\frac{V_{0}}{\pi}\right)^{\frac{1}{3}}\frac{d\theta}{dt},
\label{eq:T26}
\end{equation}
from Eq.~(\ref{eq:T16}), the evolution equation (\ref{eq:T20}) can be transformed into
\begin{equation}
\frac{1}{2}\theta^{\frac{6n+7}{3}}=\left(-\Gamma_{\rm p}\dot{\theta}\right)^{n},
\label{eq:T27}
\end{equation}
where $\dot{\theta}=d\theta/dt$, and
\begin{equation}
\Gamma_{\rm p}=\frac{2n+1}{9n}\left[2^{\frac{6n+2}{3}}k\left(\frac{V_{0}}{\pi}\right)^{\frac{1}{3}}\frac{\kappa}{\sigma_{\rm LV}}\right]^{\frac{1}{n}}
\label{eq:T28}
\end{equation}
determines the time scale of spreading on a flat substrate.  Note that the time scale $\Gamma_{\rm p}$ depends weakly on the droplet volume $V_{0}$.

Then, the solution of Eq.~(\ref{eq:T27}) is given by
\begin{equation}
\theta=\theta_{0}\left(1+\theta_{0}^{\frac{3n+7}{3n}}\frac{3n+7}{3\cdot 2^{\frac{1}{n}}n}\frac{t}{\Gamma_{\rm p}}\right)^{-\frac{3n}{3n+7}},
\label{eq:T29}
\end{equation}
where $\theta_{0}$ is the contact angle at $t=0$.  Therefore, the time evolution of the contact angle $\theta$ is asymptotically given by
\begin{equation}
\theta \propto \left(\frac{t}{\Gamma_{\rm p}}\right)^{-\frac{3n}{3n+7}},
\label{eq:T30}
\end{equation}
and the base radius $a$ expands according to
\begin{equation}
a \propto \theta^{-\frac{1}{3}} \propto  \left(\frac{t}{\Gamma_{\rm p}}\right)^{\frac{n}{3n+7}},
\label{eq:T31}
\end{equation}
and the spreading velocity $U$ decelerates according to
\begin{equation}
U \propto \dot{a} \propto  \left(\frac{t}{\Gamma_{\rm p}}\right)^{-\frac{2n+7}{3n+7}},
\label{eq:T32}
\end{equation}
which can also be derived directly from Eqs.~(\ref{eq:T21}) and (\ref{eq:T30}).  Those results for the Cassie droplet are the same as those for the free droplet on a flat smooth substrate~\cite{Liang2012}.  We can recover the well-know Tanner's law $a\propto t^{1/10}$ for Newtonian liquids from Eq.~(\ref{eq:T31}) when $n=1$.

On the other hand, the Wenzel droplet can completely wet the substrate ($\theta_{\rm e}=0^{\circ}$) as far as Eq.~(\ref{eq:T4}) is satisfied even when $\theta_{\rm Y}> 0^{\circ}$.  The evolution equation for the Wenzel droplet given by Eq.~(\ref{eq:T23}) is transformed into
\begin{equation}
\left(r_{\rm s}\cos\theta_{\rm Y}-1\right)\theta^{\frac{6n+1}{3}}=\left(-\Gamma_{\rm p}\dot{\theta}\right)^{n}
\label{eq:T33}
\end{equation}
instead of Eq.~(\ref{eq:T27}), where the factor $\left(r_{\rm s}\cos\theta_{\rm Y}-1\right)>0$ from Eq.~(\ref{eq:T4}) is the constant driving (capillary) force of spreading. The solution of Eq.~(\ref{eq:T33}) is given by
\begin{equation}
\theta=\theta_{0}\left[1+\theta_{0}^{\frac{3n+1}{3n}}\left(r\cos\theta_{\rm Y}-1\right)^{\frac{1}{n}}\frac{3n+1}{3n}\frac{t}{\Gamma_{\rm p}}\right]^{-\frac{3n}{3n+1}},
\label{eq:T34}
\end{equation}
which leads to the scaling laws
\begin{equation}
\theta \propto \left(\frac{t}{\Gamma_{\rm p}}\right)^{-\frac{3n}{3n+1}},
\label{eq:T35}
\end{equation}
and
\begin{equation}
a \propto \left(\frac{t}{\Gamma_{\rm p}}\right)^{\frac{n}{3n+1}},
\label{eq:T36}
\end{equation}
and the spreading velocity decelerates according to
\begin{equation}
U \propto  \left(\frac{t}{\Gamma_{\rm p}}\right)^{-\frac{2n+1}{3n+1}}. 
\label{eq:T37}
\end{equation}
Note that Eq.~(\ref{eq:T36}) does not reduce to the Tanner's law $a\propto t^{1/10}$ when $n=1$.  Instead, Eq.~(\ref{eq:T36}) predicts much faster spreading $a\propto t^{1/4}$ when $n=1$.  This topography driven and enhanced spreading has already been pointed out by McHale {\it et al.}~\cite{McHale2004,McHale2009} and has been confirmed experimentally~\cite{McHale2004}.  However, they studied only the Newtonian liquid with $n=1$ on a flat rough substrate.

In this paper, we are neglecting  the energy dissipation due to the imbibition within the texture of the substrate.  In fact, semi-quantitative estimation of this dissipation is possible if the substrate has a periodic ordered structure such as the artificial forest of micropillars~\cite{Ishino2007,Grewal2015}. Suppose the rough substrate is made from a periodic array of pillars with radius $b$, height $h$ and pitch (period) $p$ on a planar base, and the liquid is Newtonian.  Then, there are two dissipation channels within the texture of the substrate.  The first channel is due to the friction force at the base of the forest $F_{1}\sim \eta aU/h$ (per unit length)  ~\cite{Ishino2007}, which gives the energy dissipation 
\begin{equation}
\dot{\Sigma}_{\rm base}= 2\pi aUF_{1}\sim \frac{\eta a^{2}U^{2} }{h}.
\label{eq:T38}
\end{equation}
The second channel is due to the friction by the pillars $F_{2}\sim \eta ahU/p^{2}\ln\left(p/b\right)\sim \eta ahU/p^{2}$ (per unit length)~\cite{Hashimoto1959,Ishino2007}, which leads to 
\begin{equation}
 \dot{\Sigma}_{\rm pillar}= 2\pi aUF_{2}\sim \frac{\eta a^{2}hU^{2} }{p^{2}}.
\label{eq:T39}
\end{equation} 
The former ($\dot{\Sigma}_{\rm base}$) is dominant for short pillars ($h\ll p \ll a$), while  the latter ($\dot{\Sigma}_{\rm pillar}$) is dominant for tall pillars ($p \ll h \ll a$).

The viscous dissipation within the droplet in Eq.~(\ref{eq:T13}) is written as 
\begin{equation}
\dot{\Sigma}_{\rm drop} \sim \frac{\eta a U^{2}}{\theta},
\label{eq:T40}
\end{equation}
for Newtonian liquids with $n=1$.  Then, the dissipation at the base of the texture can be neglected when $\dot{\Sigma}_{\rm drop}\gg\dot{\Sigma}_{\rm base}$ for short pillars ($h\ll p \ll a$), which gives the condition for the dynamic contact angle $\theta$: 
\begin{equation}
\theta \ll \frac{h}{a}.
\label{eq:T41}
\end{equation}
This condition will be satisfied for the late stage of spreading when $\theta\rightarrow 0^{\circ}$. Similarly, the dissipation by the pillars of the texture can be neglected when $\dot{\Sigma}_{\rm drop}\gg\dot{\Sigma}_{\rm pillar}$ for tall pillars ($p \ll h \ll a$), which gives the condition for the dynamic contact angle: 
\begin{equation}
\theta \ll \left(\frac{p}{a}\right)\left(\frac{p}{h}\right)
\label{eq:T42}
\end{equation}
This condition is slightly more stringent than Eq.~(\ref{eq:T41}).  In these cases when Eq.~(\ref{eq:T41}) or (\ref{eq:T42}) is satisfied, we can safely neglect the energy dissipation within the texture of the substrate by imbibition.  Although, this discussion is based on the formulas for Newtonian liquids, a similar condition is expected for non-Newtonian liquids as the viscosity in $\dot{\Sigma}_{\rm drop}$, $\dot{\Sigma}_{\rm base}$ and $\dot{\Sigma}_{\rm pillar}$ will be canceled out.

\subsection{Spreading on a spherical rough substrate}

\begin{figure}[htbp]
\begin{center}
\includegraphics[width=0.6\linewidth]{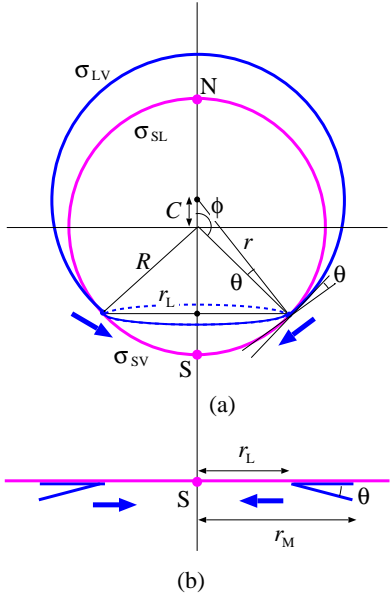}
\caption{
(a) The late stage of spreading of a spherical cap-shaped liquid droplet spreading from the north pole $N$ towards the south pole $S$ of a convex spherical substrate.  The center of the droplet with radius $r$ and that of the spherical substrate with radius $R$ are separated by a distance $C$.  The radius of the contact line is denoted by $r_{\rm L}$ and the dynamic contact angle is denoted by $\theta$, which is related to the half of the central angle $\phi$.  In the complete-wetting limit $\theta\rightarrow 0^{\circ}$ ($\phi\rightarrow 180^{\circ}$), the contact line shrinks and approaches the south pole $S$ of the substrate. 
(b) The crater model of late stage of spreading droplet on a spherical substrate~\cite{Iwamatsu2017b}.  The spreading droplet is modeled by a shrinking crater with a wedge-shaped meniscus, which is characterized by its length $r_{\rm M}$ and the radius of the crater $r_{\rm L}$.  The three-phase contact line shrinks towards the south pole $S$ of the substrate. 
 }
\label{fig:T4}
\end{center}
\end{figure}

To study the spreading on a spherical rough substrate, we concentrate on the late-stage of spreading shown in Fig.~\ref{fig:T4}(a).  We model the spreading of the three-phase contact line of cap-shaped droplet with radius $r$ towards the south pole $S$ of the spherical substrate with radius $R$ as a shrinking crater on a flat substrate (Fig.~\ref{fig:T4}(b))~\cite{Iwamatsu2017b}.  We adopt the energy balance approach of hydrodynamic model~\cite{deGennes1985,Liang2012,Iwamatsu2017b} again.  Then the viscous energy dissipation is balanced by the work done by the driving capillary force of spreading given by
\begin{equation}
f_{\rm L}=\sigma_{\rm LV}\left(\cos\theta_{\rm e}-\cos\theta\right) - \frac{\tau}{R\tan\phi},
\label{eq:T43}
\end{equation}
for the droplet on a spherical rough substrate, where $\theta_{\rm e}$ is the equilibrium contact angle, $\phi$ is a half of the central angle (Fig.~\ref{fig:T4}(a)).  Again, we will neglect the line tension $\tau$ in the following discussion.  Then, the capillary force on a spherical substrate in Eq.~(\ref{eq:T43}) becomes identical to those on a flat substrate given by Eq.~(\ref{eq:T19}) for the Cassie droplet and Eq.~(\ref{eq:T22}) for the Wenzel droplet.

Since we neglect the liquid volume within the textured substrate, we will only consider the viscous dissipation $\dot{\Sigma}_{\rm drop}$ within the cap-shaped spherical droplet again.  Now, the spreading droplet on a spherical substrate can be modeled by a shrinking crater with a wedge-shaped meniscus as shown in Fig.~\ref{fig:T4}(b).  Then the viscous dissipation $\dot{\Sigma}_{\rm drop}$ within the crater is given by~\cite{Iwamatsu2017b}
\begin{equation}
\dot{\Sigma}_{\rm drop}=2\pi\lambda\left(\frac{2n+1}{n}\right)^{n}\frac{\kappa U^{n+1}}{\theta^{n}}r_{\rm L}^{2-n},
\label{eq:T44}
\end{equation}
where $r_{\rm L}=R\sin\phi$ is the radius of the three-phase contact line.  The coefficient $\lambda$ is determined from the cut-off length for $r_{\rm L}$ and the length $r_{\rm M}$ (Fig.~\ref{fig:T4}(b)) of wedge~\cite{Iwamatsu2017b}.

The energy-balance condition at the contact line L with radius $r_{\rm L}$
\begin{equation}
2\pi r_{\rm L} f_{\rm L}U=\dot{\Sigma}_{\rm drop}
\label{eq:T45}
\end{equation}
is given by
\begin{equation}
\theta^{n}\left(\cos\theta_{\rm e}-\cos\theta\right)=\lambda\left(\frac{2n+1}{n}\right)^{n}r_{\rm L}^{1-n}\frac{\kappa U^{n}}{\sigma_{\rm LV}},
\label{eq:T46}
\end{equation}
which is similar to Eq.~(\ref{eq:T15}).

From the geometrical relation of a cap-shaped droplet on a spherical substrate given by
\begin{equation}
\tan\phi=\frac{r\sin\theta}{R-r\cos\theta},
\label{eq:T47}
\end{equation}
we can obtain~\cite{Iwamatsu2017b}
\begin{equation}
r_{\rm L}=R\sin\phi\simeq \frac{r_{0}}{r_{0}-R}R\theta,
\label{eq:T48}
\end{equation}
as $\theta\rightarrow 0^{\circ}$ where $r_{0}$ is the droplet radius when it completely encloses the spherical substrate, and it is determined from the droplet volume $V_{0}$ through
\begin{equation}
V_{0}=\frac{4\pi}{3}\left(r_{0}^{3}-R^{3}\right).
\label{eq:T49}
\end{equation}
Then, the energy-balance condition in Eq.~(\ref{eq:T46}) can be written as
\begin{equation}
\theta^{2n-1}\left(\cos\theta_{\rm e}-\cos\theta\right)
=\lambda\left(\frac{2n+1}{n}\right)^{n}\left(\frac{r_{0}}{r_{0}-R}R\right)^{1-n}\frac{\kappa U^{n}}{\sigma_{\rm LV}},
\label{eq:T50}
\end{equation}
where the equilibrium contact angle $\theta_{\rm e}$ is given either by Eq.~(\ref{eq:T1}) for the Wenzel droplet or by Eq.~(\ref{eq:T9}) for the Cassie droplet.  The difference of exponent of $\theta$ in Eq.~(\ref{eq:T50}) for a spherical substrate from that in Eq.~(\ref{eq:T17}) for a flat substrate originates from the fact that the base radius of droplet on a spherical substrate $r_{\rm L}$ shrinks according to $r_{\rm L}\propto \theta$ from Eq.~(\ref{eq:T48}) while the base radius $a$ on a flat substrate expands according to $a\propto \theta^{-1/3}$ from Eq.~(\ref{eq:T16}).  Therefore, this difference comes from the topology (flat and spherical geometry).

When the equilibrium contact angle $\theta_{\rm e}$ is given by the Cassie-Baxter formula in Eq.~(\ref{eq:T9}) and the capillary force for spreading is given by Eq.~(\ref{eq:T19}), Eq.~(\ref{eq:T50}) for the Cassie droplet is further simplified to
\begin{equation}
\frac{1}{2}\theta^{2n-1}\left(\theta^{2}-\phi_{\rm s}\theta_{\rm Y}^{2}\right)
=\lambda\left(\frac{2n+1}{n}\right)^{n}\left(\frac{r_{0}}{r_{0}-R}R\right)^{1-n}\frac{\kappa U^{n}}{\sigma_{\rm LV}},
\label{eq:T51}
\end{equation}
when both $\theta$ and $\theta_{\rm Y}$ are small.  Equation (\ref{eq:T51}) leads to the relationship between the dynamic contact angle $\theta$ and the spreading velocity $U$ given by
\begin{equation}
\theta^{2n+1}\propto U^{n},
\label{eq:T52}
\end{equation} 
for the complete wetting ($\theta_{\rm Y}=0^{\circ}$) substrate.  This formula for the Cassie droplet is similar to those derived for the Cassie and the free droplet on a flat smooth substrate~\cite{Liang2012,Wang2007} given in Eq.~(\ref{eq:T21}), though the exponent is different. This difference originates from the difference of Eq.~(\ref{eq:T50}) and Eq.~(\ref{eq:T17}), which comes from the  topology of the substrate.   All those results~\cite{Liang2012,Wang2007}, including our Eq.~(\ref{eq:T52}) reduce to the universal law given by Eq.~(\ref{eq:T18}) for the Newtonian liquids ($n=1$).

When the equilibrium contact angle is given by the Wenzel formula in Eq.~(\ref{eq:T1}), the capillary force is given by Eq.~(\ref{eq:T22}), and Eq.~(\ref{eq:T50}) for the Wenzel droplet is further simplified to
\begin{eqnarray}
&&\theta^{2n-1}\left[r_{\rm s}\cos\theta_{\rm Y}-1+\frac{1}{2}\theta^{2}\right] 
\nonumber \\
&&=\lambda\left(\frac{2n+1}{n}\right)^{n}\left(\frac{r_{0}}{r_{0}-R}R\right)^{1-n}\frac{\kappa U^{n}}{\sigma_{\rm LV}}.
\label{eq:T53}
\end{eqnarray}
This relationship between the dynamic contact angle $\theta$ and the spreading velocity $U$ given by
\begin{equation}
\theta^{2n-1}\propto U^{n},
\label{eq:T54}
\end{equation}
for a complete wetting substrate characterized by Eq.~(\ref{eq:T4}) is different from that derived for the free droplet on a spherical smooth substrate~\cite{Iwamatsu2017b}.  Equation (\ref{eq:T54}) reduces to Eq.~(\ref{eq:T25}) for the droplet on a flat substrate when the liquid is Newtonian ($n=1$).  Since Equation (\ref{eq:T54}) is different from Eq.~(\ref{eq:T24}), the scaling laws for the Wenzel droplet will be different from those for the Cassie droplets not only on a flat but also on a spherical substrate. 

When the contact angle $\theta$ is low, the spreading velocity $U$ on a spherical substrate is given by~\cite{Iwamatsu2017b}
\begin{equation}
U=\frac{d}{dt}R\phi=-\frac{r_{0}}{r_{0}-R}R\dot{\theta}
\label{eq:T55}
\end{equation}
from Eq.~(\ref{eq:T48}), where $\dot{\theta}<0$. Again, Equation (\ref{eq:T55}) is different from Eq.~(\ref{eq:T26}) by the difference of the (flat and spherical) topology of the substrate.  Then, Eq.~(\ref{eq:T51}) for the Cassie droplet is written as
\begin{equation}
\frac{1}{2}\theta^{2n+1}
=\left(-\Gamma_{\rm s}\dot{\theta}\right)^{n},
\label{eq:T56} 
\end{equation}
with
\begin{equation}
\Gamma_{\rm s}=\left[\lambda\left(\frac{r_{0}}{r_{0}-R}\right)\frac{\kappa}{\sigma_{\rm LV}}\right]^{\frac{1}{n}}\left(\frac{2n+1}{n}\right)R^{\frac{1}{n}},
\label{eq:T57}
\end{equation}
which determines the time scale of spreading on a completely wettable spherical substrate ($\theta_{\rm Y}=0^{\circ}$).  Note that the time scale $\Gamma_{\rm s}$ depends on the power exponent $n$, the radius of the substrate $R$ and the volume of the droplet $V_{0}$ through $r_{0}$, and, in particular, $\Gamma_{\rm s}$ is proportional to $R^{1/n}$ of the substrate $\Gamma_{\rm s}\propto R^{1/n}$.  The time evolution equation in Eq.~(\ref{eq:T56}) and the time scale in Eq.~(\ref{eq:T57}) of spreading on a spherical substrate is the same as that derived for the free droplet on a spherical smooth substrates~\cite{Iwamatsu2017b}.

When the substrate is completely wettable ($\theta_{\rm Y}=0^{\circ}$), we can solve Eq.~(\ref{eq:T56}) in the limit $\theta\rightarrow 0^{\circ}$ and obtain the time evolution of the contact angle of the Cassie droplet which is the same as that on a smooth spherical substrate~\cite{Iwamatsu2017b}
\begin{equation}
\theta = \theta_{0}\left(1+\theta_{0}^{\frac{n+1}{n}}\frac{n+1}{2^{\frac{1}{n}}n}\left(\frac{t}{\Gamma_{\rm s}}\right)\right)^{-\frac{n}{n+1}},
\label{eq:T58}
\end{equation}
where $\theta_{0}$ is the contact angle at $t=0$.  Therefore, the time evolution of the contact angle is asymptotically given by
\begin{equation}
\theta \propto \left(\frac{t}{\Gamma_{\rm s}}\right)^{-\frac{n}{n+1}},
\label{eq:T59}
\end{equation}
and the radius $r_{\rm L}$ of the contact circle shrinks according to
\begin{equation}
r_{\rm L} \propto \theta \propto \left(\frac{t}{\Gamma_{\rm s}}\right)^{-\frac{n}{n+1}},
\label{eq:T60}
\end{equation}
whose spreading velocity decelerates according to
\begin{equation}
U \propto -\dot{\theta} \propto \left(\frac{t}{\Gamma_{\rm s}}\right)^{-\frac{2n+1}{n+1}},
\label{eq:T61}
\end{equation}
from Eq.~(\ref{eq:T58}) and (\ref{eq:T52}).  They are the same as those for the free droplet on a spherical smooth substrate~\cite{Iwamatsu2017b}, but are different from those for the Cassie droplet or the free droplet on a smooth flat substrate~\cite{Liang2012} given by Eqs.~(\ref{eq:T30})-(\ref{eq:T32}).

The evolution equation in Eq.~(\ref{eq:T53}) for the Wenzel droplet is simplified to 
\begin{equation}
\theta^{2n-1}\left(r_{\rm s}\cos\theta_{\rm Y}-1\right)
=\left(-\Gamma_{\rm s}\dot{\theta}\right)^{n},
\label{eq:T62}
\end{equation}
whose solutions are different for shear-thickening liquids with $n>1$, shear-thinning liquids with $n<1$ and  Newtonian liquids with $n=1$.  For Newtonian liquids with $n=1$, Eq.~(\ref{eq:T62}) becomes
\begin{equation}
\theta \left(r_{\rm s}\cos\theta_{\rm Y}-1\right)=\left(-\Gamma_{\rm s}\dot{\theta}\right),
\label{eq:T63}
\end{equation}
whose solution is given by
\begin{equation}
\theta = \theta_{0}\exp\left[-\left(r_{\rm s}\cos\theta_{\rm Y}-1\right)\left(\frac{t}{\Gamma_{\rm s}}\right)\right].
\label{eq:T64}
\end{equation}
Therefore, all physical quantities decay exponentially as
\begin{equation}
\theta \propto r_{\rm L} \propto U \propto \exp\left(-t/t_{\rm s}\right),
\label{eq:T65}
\end{equation}
where
\begin{equation}
t_{\rm s}=\frac{\Gamma_{\rm s}}{r_{\rm s}\cos\theta_{\rm Y}-1}
\label{eq:T66}
\end{equation}
is the decay constant of the exponential relaxation.

The solution of Eq.~(\ref{eq:T62}) for shear-thickening liquids with $n>1$ is given by
\begin{equation}
\theta=\theta_{0}\left[1+\theta_{0}^{\frac{n-1}{n}}\frac{n-1}{n}\left(r_{\rm s}\cos\theta_{\rm Y}-1\right)^{\frac{1}{n}}\left(\frac{t}{\Gamma_{\rm s}}\right)\right]^{-\frac{n}{n-1}},
\label{eq:T67}
\end{equation}
which gives the asymptotic scaling law for the contact angle
\begin{equation}
\theta \propto \left(\frac{t}{\Gamma_{\rm s}}\right)^{-\frac{n}{n-1}},
\label{eq:T68}
\end{equation}
and that for the contact line radius
\begin{equation}
r_{\rm L} \propto \left(\frac{t}{\Gamma_{\rm s}}\right)^{-\frac{n}{n-1}}.
\label{eq:T69}
\end{equation}
Then, the spreading velocity decelerates according to
\begin{equation}
U \propto \left(\frac{t}{\Gamma_{\rm s}}\right)^{-\frac{2n-1}{n-1}}.
\label{eq:T70}
\end{equation}

In contrast to the shear-thickening liquid with $n>1$, the spreading of the shear-thinning liquid with $n<1$ is faster and finishes within a finite time $t_{0}$.  The solution of Eq.~(\ref{eq:T62}) is given by
\begin{equation}
\theta=\left[\frac{1-n}{n}\left(r_{\rm s}\cos\theta_{\rm Y}-1\right)^{\frac{1}{n}}\right]^{\frac{n}{1-n}}\left(\frac{t_{0}-t}{\Gamma_{\rm s}}\right)^{\frac{n}{1-n}},
\label{eq:T71}
\end{equation}
where the spreading time $t_{0}$ can be related to the initial contact angle $\theta_{0}$ at $t=0$.
Therefore, the contact angle and the radius of the contact line vanish within a finite time $t_{0}$ according to
\begin{equation}
\theta \propto \left(\frac{t_{0}-t}{\Gamma_{\rm s}}\right)^{\frac{n}{1-n}},
\label{eq:T72}
\end{equation}
and
\begin{equation}
r_{\rm L} \propto \left(\frac{t_{0}-t}{\Gamma_{\rm s}}\right)^{\frac{n}{1-n}},
\label{eq:T73}
\end{equation}
and the spreading velocity changes according to
\begin{equation}
U \propto \left(\frac{t_{0}-t}{\Gamma_{\rm s}}\right)^{\frac{2n-1}{1-n}}.
\label{eq:T74}
\end{equation}
The spreading velocity $U$ decelerates and approaches zero when $1>n>1/2$, and it accelerates and diverges as $t\rightarrow t_{0}$ when $n<1/2$.  The spreading velocity $U$ remains constant up to the terminal time $t_{0}$ when $n=1/2$.  Thus, the shear-thinning liquid with $n<1$ does not follow the power-law scaling rule.  Such an anomalous asymptotic behavior has also been predicted on smooth spherical substrates when the line tension plays a dominant role~\cite{Iwamatsu2017,Iwamatsu2017b}.

\subsection{Scaling laws of topography and topology driven spreading}

In Table~\ref{tab:T1}, we summarize various spreading laws on a flat and a spherical rough substrate derived in this paper. The results for the Cassie droplet are the same as those for the free droplet on a smooth substrate.  Therefore, the results for the Cassie droplet reduce to the Tanner's law~\cite{Tanner1979} for Newtonian liquids when $n=1$.  Most notably, the spreading of the Wenzel droplet of shear-thinning liquids with $n<1$ on a spherical rough substrate is the fastest.  In fact, the spreading finishes within a finite time.  Almost the same fast spreading is predicted on a smooth spherical substrate when the spreading is driven by the line tension~\cite{Iwamatsu2017b}.  On a spherical substrate, the contact line shrinks as the spreading proceeds, and the positive line tension accelerate the spreading toward the complete wetting~\cite{Iwamatsu2017,Iwamatsu2017b}.  Even without the line tension, the roughness of substrate always provides capillary driving force in Eq.~(\ref{eq:T22}) for the Wenzel droplet and accelerates the spreading, which was already noted by McHale et al.~\cite{McHale2004,McHale2009}.  On a spherical substrate, this {\it topography} driven spreading~\cite{McHale2004} is further enhanced by the difference of {\it topology} of the substrate, which leads to the {\it topography and topology driven spreading}.  So, the topography and the topology synergetically accelerate the spreading toward the complete wetting state on a spherical rough substrate.  

\begin{table}[htb]
 \begin{center}
  \caption{Various spreading laws on a flat and a spherical substrate}
  \begin{tabular}{crrrr} 
\hline
 & \multicolumn{2}{c}{Flat substrate} & \multicolumn{2}{c}{Spherical substrate}   \\
\hline
&Cassie & Wenzel & Cassie & Wenzel \\
\hline
$U^{n}$ &
$\theta^{\frac{2n+7}{3}}$ & 
$\theta^{\frac{2n+1}{3}}$ & 
$\theta^{2n+1}$ & 
$\theta^{2n-1}$ \\
$\theta$ &
$t^{-\frac{3n}{3n+7}}$ &
$t^{-\frac{3n}{3n+1}}$ &
$t^{-\frac{n}{n+1}}$ &
$t^{-\frac{n}{n-1}}$ ($n>1$)\\
&&&&$\exp\left(-t/t_{\rm s}\right)$ ($n=1$) \\
&&&&$\left(t_{0}-t\right)^{\frac{n}{1-n}}$ ($n<1$)\\
$a$, $r_{\rm L}$ &
$t^{\frac{n}{3n+7}}$ &
$t^{\frac{n}{3n+1}}$ &
$t^{-\frac{n}{n+1}}$ &
$t^{-\frac{n}{n-1}}$ ($n>1$)\\
&&&&$\exp\left(-t/t_{\rm s}\right)$ ($n=1$) \\
&&&&$\left(t_{0}-t\right)^{\frac{n}{1-n}}$ ($n<1$)\\
$U$ &
$t^{-\frac{2n+7}{3n+7}}$ &
$t^{-\frac{2n+1}{3n+1}}$ &
$t^{-\frac{2n+1}{n+1}}$ &
$t^{-\frac{2n-1}{n-1}}$ ($n>1$)\\
&&&&$\exp\left(-t/t_{\rm s}\right)$ ($n=1$) \\
&&&&$\left(t_{0}-t\right)^{\frac{2n-1}{1-n}}$ ($n<1$)\\
\hline
  \end{tabular}
  \label{tab:T1}
 \end{center}
\end{table}

\begin{figure}[htbp]
\begin{center}
\includegraphics[width=0.90\linewidth]{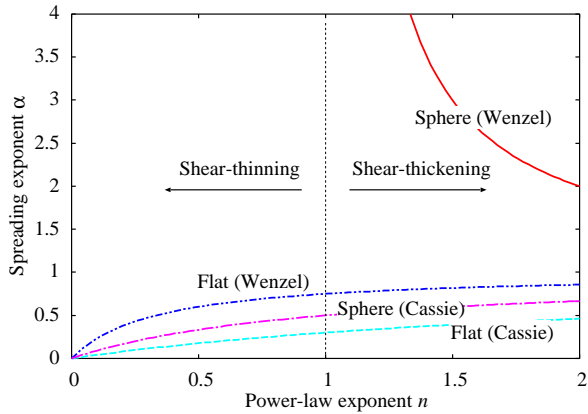}
\caption{
The spreading exponent $\alpha$ of the dynamic contact angle $\theta\propto t^{-\alpha}$ summarize in Tab.~\ref{tab:T1} for the Cassie and the Wenzel droplet on a flat and a spherical rough substrates as a function of the power-law exponent $n$ of non-Newtonian liquids. The evolution of the Wenzel droplet of non-Newtonian shear-thinning liquids with $n<1$ and that of Newtonian liquids with $n=1$ are not described by the power law relaxation. 
}
\label{fig:T5}
\end{center}
\end{figure}

In Fig.~\ref{fig:T5}, we compare the spreading exponent $\alpha$ of the dynamic contact angle defined by
\begin{equation}
\theta \propto t^{-\alpha}
\label{eq:T75}
\end{equation}
for the Cassie and the Wenzel droplet on a flat and a spherical rough substrate except for the shear-thinning ($n<1$) Wenzel droplet on a spherical substrate.  They are given by  $\alpha_{\rm flat\left(Cassie\right)}=3n/\left(3n+7\right)$, $\alpha_{\rm flat\left(Wenzel\right)}=3n/\left(3n+1\right)$, $\alpha_{\rm sphere\left(Cassie\right)}=n/\left(n+1\right)$, and $\alpha_{\rm sphere\left(Wenzel\right)}=n/\left(n-1\right)$ as functions of the power-law exponent $n$.  Generally, the following inequality
\begin{equation}
\alpha_{\rm sphere\left(Wenzel\right)}>\alpha_{\rm flat\left(Wenzel\right)}>\alpha_{\rm sphere\left(Cassie\right)}>\alpha_{\rm flat\left(Cassie\right)},
\label{eq:T76}
\end{equation}
holds for the same power-law exponent $n$.  The spreading of a Wenzel droplet  is faster than that of a Cassie droplet on the same type of substrate ($\alpha_{\rm sphere\left(Wenzel\right)}>\alpha_{\rm sphere\left(Cassie\right)}$ and $\alpha_{\rm flat\left(Wenzel\right)}>\alpha_{\rm flat\left(Cassie\right)}$). The spreading on a spherical substrate is faster than that on a flat substrate for the same type of droplet ($\alpha_{\rm sphere\left(Wenzel\right)}>\alpha_{\rm flat\left(Wenzel\right)}$ and $\alpha_{\rm sphere\left(Cassie\right)}>\alpha_{\rm flat\left(Cassie\right)}$), which is due to the efficiency of energy-dissipation.  In fact, the radius of the contact line $r_{\rm L}$ shrinks on a spherical substrate, while the radius $a$ expands infinity on a flat substrate.  Therefore, the energy dissipation by viscosity is less effective on a flat substrate than on a spherical substrate, which leads to a weaker braking force on a flat substrate.

The spreading exponent $\alpha$ of the Cassie and the Wenzel droplet on a flat substrate, and that of the Cassie droplet on a spherical substrate are increasing functions of the power-law exponent $n$.  In contrast, the spreading exponent of the Wenzel droplet of shear thickening liquids with $n>1$ is a decreasing function.  Furthermore, it diverges as $n\rightarrow 1^{+}$.  Therefore, the spreading of the Wenzel droplet with $n>1$ on a spherical substrate is much faster than the others.

\begin{figure}[htbp]
\begin{center}
\subfigure[]
{
\includegraphics[width=0.45\linewidth]{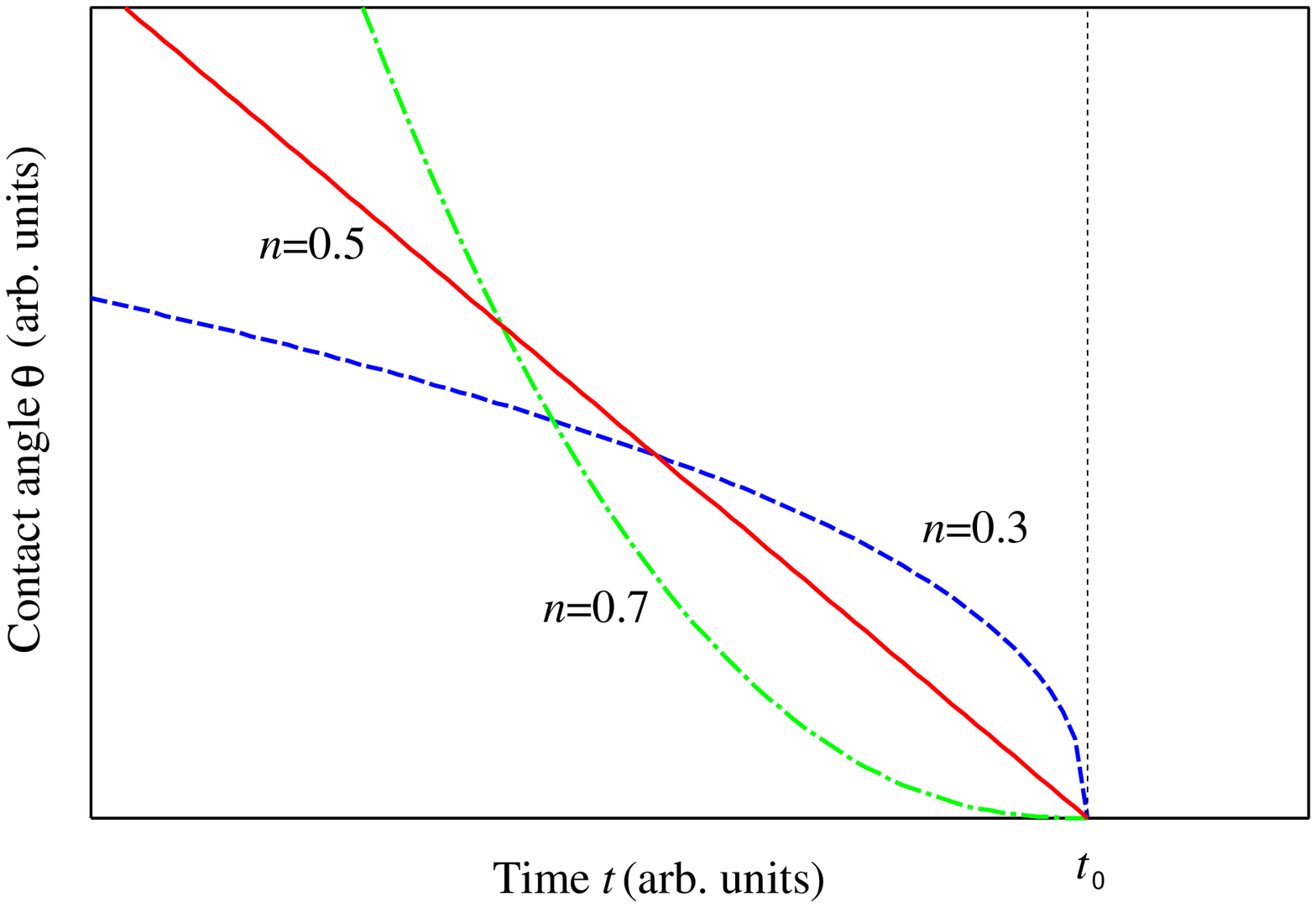}
\label{fig:T6a}
}
\subfigure[]
{
\includegraphics[width=0.45\linewidth]{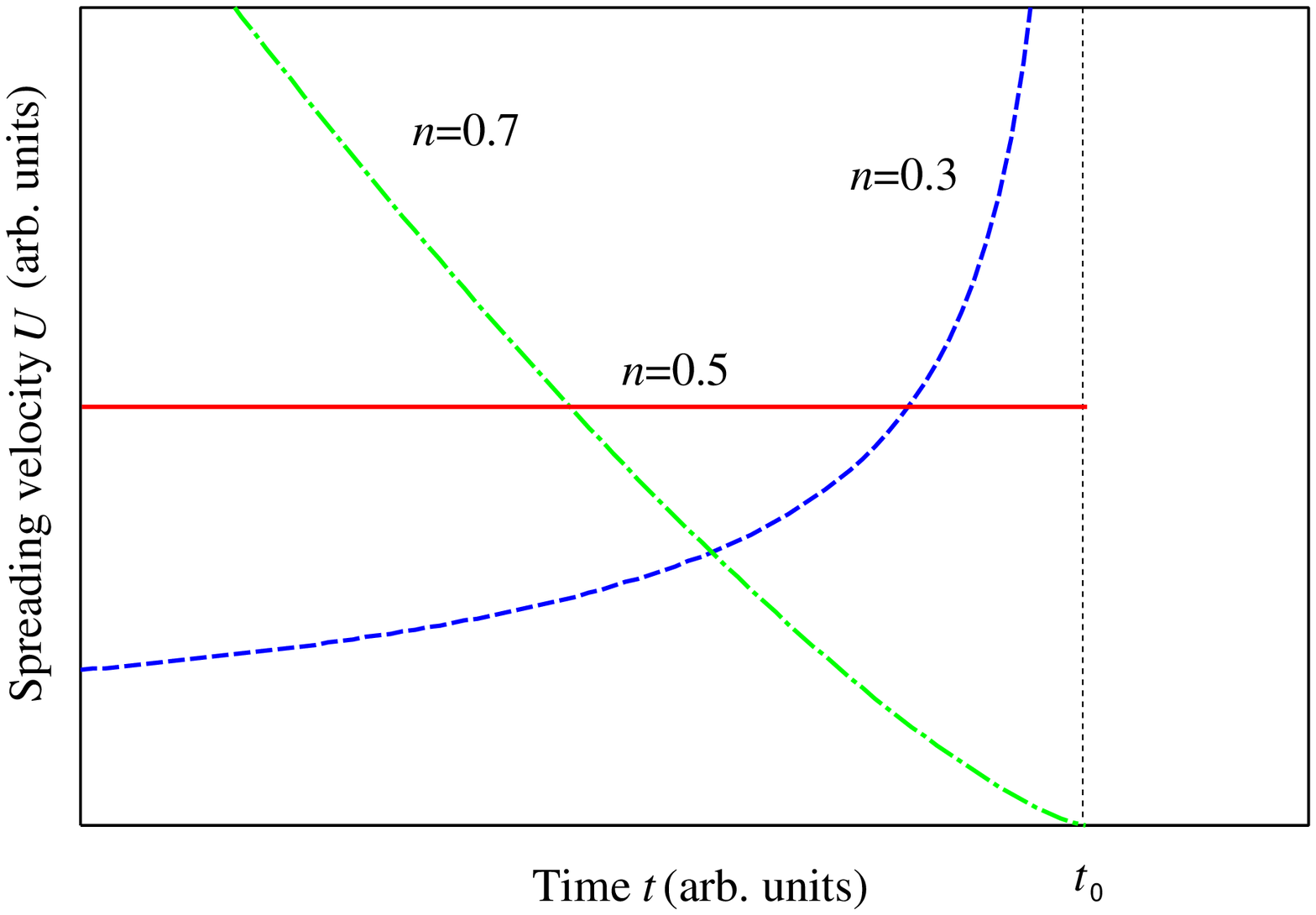}
\label{fig:T6b}
}
\end{center}
\caption{
(a) The evolution of the dynamic contact angle $\theta$ of the Wenzel droplet of non-Newtonian shear-thinning liquids with $n<1$ on a spherical rough substrate as a function of time $t$.  The contact angle vanishes at the terminal time $t_{0}$.  When $n=0.5$, the contact angle decreases linearly with time.  The contact angle decreases gradually when $n>0.5$, while it decreases suddenly near the terminal time $t_{0}$ when $n<0.5$.  (b) The spreading velocity $U\propto -\dot{\theta}$ as a function of time. The velocity is constant when $n=0.5$.  It accelerates and diverges at $t_{0}$ when $n<0.5$ , while it decelerates and vanishes at $t_{0}$ when $n>0.5$.
 } 
\label{fig:T6}
\end{figure}

The spreading of the Wenzel droplet of Newtonian ($n=1$) and non-Newtonian shear-thinning ($n<1$) liquids on a spherical substrate does not follow the power-law relaxation (Table~\ref{tab:T1}).  In contrast to the power-law spreading, which is commonly observed in various Newtonian ($n=1$)~\cite{Tanner1979,deGennes1985} as well as non-Newtonian ($n\neq 1$) liquid droplets~\cite{Liang2012,Iwamatsu2017}, the spreading of the Wenzel droplet of Newtonian liquid ($n=1$) is exponential (Table~\ref{tab:T1}) and, therefore, is faster than the power-law spreading.

The spreading time of the Wenzel droplet of non-Newtonian shear-thinning liquids with $n<1$ becomes even finite.  Furthermore, the spreading behavior is divided into three categories according to the magnitude of the power-law exponent $n$ as shown in Fig.~\ref{fig:T6}.  The contact angle $\theta$ approaches $0^{\circ}$ linearly when $n=0.5$.  It decreases gradually when $n>0.5$ and it decreases suddenly near the terminal time $t_{0}$ when $n<0.5$ (Fig.~\ref{fig:T6}(a)).   Correspondingly, the spreading velocity $U$ is constant when $n=0.5$, and it decelerates and approaches $0$ as $t\rightarrow t_{0}$ when $n>0.5$.  In contrast, the spreading velocity accelerates and diverges $U\rightarrow \infty$ as $t\rightarrow t_{0}$ when $n<0.5$ (Fig.~\ref{fig:T6}(b))

In addition to the {\it topology driven wetting}~\cite{Iwamatsu2016a,Iwamatsu2016b} on a spherical substrate, all those results summarized in Table~\ref{tab:T1} reveal the diversity of  {\it topography and topology driven spreading} behavior of cap-shaped droplets on a flat and a spherical rough substrate.  

Although there exists a wealth of experimental data of spreading of Newtonian as well as non-Newtonian liquids on a flat smooth substrate~\cite{Bonn2009,Liang2009,Liang2012} the experimental data on a flat rough substrate are scarce~\cite{Oliver1980,Apel-Paz1999,McHale2004,Xu2008,McHale2009}. Furthermore, previous experimental works on a flat rough substrate pay most attention to imbibition~\cite{Neogi1983,Daniel2006,Ishino2007,Haidara2008,Grewal2015} rather than spreading. The problem of spreading on a spherical substrate has attracted almost no attention so far except for a few experimental attempts on a smooth~\cite{Tao2011,Guilizzoni2011,Eral2011,Extrand2012} and a rough spherical substrate~\cite{Byon2010}.  For example, Byon et al.~\cite{Byon2010} used nanostructured superhydrophilic Cu spheres whose diameter is about 1 to 2 mm, and studied the drag reduction.  It will be possible to use similar nanostructured spheres to study the wetting and spreading using various types of shear-thinning and shear-thickening non-Newtonian liquids~\cite{Liang2009,Liang2012}. Then, it will be possible to confirm the various scenario of topography and topolpgy driven spreading presented in Tab.~\ref{tab:T1}.  

\section{\label{sec:sec4} Conclusion}

In the present study, we considered the problem of spreading of a cap-shaped spherical droplet of non-Newtonian liquids on a flat and a spherical rough substrate using the energy balance approach.  The Wenzel~\cite{Wenzel1936} and the Cassie-Baxter~\cite{Cassie1944} model are adopted to define the equilibrium contact angle on a rough substrate.  Only the viscous dissipation within the droplet is considered, and the dissipation within the texture of the rough substrate~\cite{Ishino2007,Haidara2008,Grewal2015} is neglected.  The viscous dissipation is calculated using the cone-shaped model~\cite{McHale2009,Liang2012} on a flat substrate and the crater-shaped model~\cite{Iwamatsu2017b} on a spherical substrate.  We derive various scaling laws for the time evolution of the dynamic contact angle similar to the Tanner's law~\cite{Tanner1979} when the substrate is completely wettable.  

The scaling exponents of the Wenzel model are different from those of the Cassie-Baxter model by the difference of {\it topography} of the substrate.  Likewise, the exponents on a flat and those on a spherical substrate are different by the difference of {\it topology} of the substrate.  The scaling exponents are determined from the power-law exponent of non-Newtonian liquids.  The spreading of Newtonian liquids and non-Newtonian shear-thinning liquids of the Wenzel model on a spherical substrate are exceptional, and does not follow the power-law.  Instead, the evolution of the contact angle of Newtonian liquids follows the exponential relaxation, and that of non-Newtonian shear-thinning liquids finishes within a finite time. The spreading of the Wenzel droplet is faster than that of the Cassie droplet on the same flat or spherical substrate due to the topography of the substrate.  Similarly, the spreading on a spherical substrate is faster than that on a flat substrate for the Wenzel or the Cassie droplet due to the topology of the substrate. The {\it topography and topology} driven spreading is fastest for the Wenzel droplet of shear-thinning liquid on a spherical substrate.  Although we did not include the line tension effect since its nature on rough substrates is largely unknown and would be the issue of debate, it is easy to inclusion the line  tension effect into our present formulation~\cite{Law2017,Iwamatsu2017,Iwamatsu2017b}.  The inclusion of line tension will further increase the variety of the topography and topology driven spreading scenario.  

There has been a very few experiment of wetting and spreading on a flat rough substrate~\cite{Daniel2006,Oliver1980,Apel-Paz1999,McHale2004,Xu2008}.  The number of experimental work of wetting and spreading on a spherical smooth substrate~\cite{Iwamatsu2016a,Iwamatsu2016b,Iwamatsu2017,Iwamatsu2017b} is very limited~\cite{Tao2011,Eral2011,Guilizzoni2011,Extrand2012}.  There seems no experimental work of spreading on a spherical rough substrate.  The experimental verification of our theoretical predictions is highly desired and urgent.  The theoretical results presented in this paper together with those future experimental results would be valuable to design and develop new nano-materials, nano-devices and new engineering applications.

\begin{acknowledgments}
This work was partially supported under a project for strategic advancement of research infrastructure for private universities, 2015-2020, operated by MEXT, Japan. A part of this work was conducted as a visiting scientist of the Department of Physics, Tokyo Metropolitan University.  The author is grateful to Professor Hiroyuki Mori and Professor Yutaka Okabe for continuous support and encouragement.  The author is also grateful to Professor Siegfried Dietrich (Max-Planck Institute for Intelligent Systems, Stuttgart) for sending him a useful material on line tension.
\end{acknowledgments}




\end{document}